\documentclass[9pt,technote]{IEEEtran}
\usepackage{lscape,url}
\usepackage{cite}
\usepackage{latexsym}
\usepackage{amsmath}
\usepackage{amstext,amsfonts,epsf,epsfig,pifont}
\usepackage{graphics,graphicx,bm,amsmath}

\newlength{\intwidth}

\def\XXint#1#2#3{{\setbox0=\hbox{$#1{#2#3}{\int}$}
\vcenter{\hbox{$#2#3$}}\kern-.5\wd0}}

\newcommand{\sinc}{\ensuremath{\mathrm{sinc}}}

\begin{document}
\noindent \emph{The following statements are placed here in accordance with the copyright policy of the Institute of Electrical and Electronics Engineers, Inc., available online at}
\texttt{http://www.ieee.org/publications\!\_standards}\\\texttt{/publications/rights/rights\!\_policies.html}\\

\noindent
Lilly, J. M., \&  Olhede, S. C. (2012).  Generalized Morse wavelets as \\\indent a superfamily of analytic wavelets.  Submitted to \emph{IEEE Transactions}\\\indent \emph{on Signal Processing}.\\


\noindent \copyright 2012 IEEE. Personal use of this material is permitted. Permission from IEEE must be obtained for all other uses, in any current or future media, including reprinting/republishing this material for advertising or promotional purposes, creating new collective works, for resale or redistribution to servers or lists, or reuse of any copyrighted component of this work in other works.\\

\newpage

$\,$\vspace{10in}
\newpage

\title{Generalized Morse Wavelets \\ as a Superfamily of Analytic Wavelets}
\author{Jonathan~M.~Lilly and Sofia~C.~Olhede
\thanks{Manuscript submitted \today.  The work of J. M. Lilly was supported  by award \#0849371 from the Physical Oceanography program of the United States National Science Foundation.  The work of S. C. Olhede was supported by award \#EP/I005250/1 from the Engineering and Physical Sciences Research Council of the United Kingdom. }
\thanks{J.~M.~Lilly is with NorthWest Research Associates, PO Box 3027, Bellevue, WA, USA (e-mail: lilly@nwra.com).}
\thanks{S.~C.~Olhede is with the Department of Statistical Science, University College London, Gower Street,
London WC1E 6BT, UK (e-mail: s.olhede@ucl.ac.uk).}}

\markboth{IEEE Transactions on Signal Processing}{Lilly \& Olhede: Analytic Wavelets}

\maketitle

\begin{abstract}
The generalized Morse wavelets are shown to constitute a superfamily that essentially encompasses all other commonly used analytic wavelets, subsuming eight apparently distinct types of analysis filters into a single common form.  This superfamily of analytic wavelets provides a framework for systematically investigating wavelet suitability for various applications.  In addition to a parameter controlling the time-domain duration or  Fourier-domain bandwidth, the wavelet {\em shape} with fixed bandwidth may be modified by varying a second parameter, called~$\gamma$.  For integer values of~$\gamma$, the most symmetric, most nearly Gaussian,  and  generally most time-frequency concentrated member of the superfamily is found to occur for $\gamma=3$.  These wavelets, known as ``Airy wavelets,'' capture the essential idea of popular Morlet wavelet, while avoiding its deficiencies.  They may be recommended as an ideal starting point for general purpose use.
\end{abstract}

\IEEEpeerreviewmaketitle

\section{Introduction}

Analytic wavelets---complex-valued time/frequency localized filters with vanishing support on negative frequencies---provide the basis for a powerful analysis of oscillatory signals, see e.g. \cite{lilly10-itit} and references therein.  In order to study time- and/or frequency-localized variability of a signal $x(t)$, the continuous wavelet transform
  \begin{equation}
\label{vectorwavetrans}
W_{\psi}(t ,s)  \equiv  \int_{-\infty}^{\infty} \frac{1}{s^n}\,
  \psi^*\left(\frac{\tau-t}{s}\right)\,x(\tau)\,d \tau
\end{equation}
is formed. This is seen as a set of bandpass operations with the scale normalization $n=1$, or as a set of projections with the normalization $n=1/2$, indexed by the scale parameter $s$.

A practical question is which wavelet to use and why.  In addition to the popular but only approximately analytic Morlet wavelet, a variety of analytic wavelets have been proposed, including the Cauchy-Klauder-Morse-Paul, Derivative of Gaussian,  lognormal or log Gabor, Shannon, and Bessel wavelets \cite{holschneider,mallat,knutsson94-itip,field87-josaa}.   However, this profusion of wavelet types serves as a barrier to acquiring the knowledge necessary to carry out practical applications.    In the absence of a suitable unifying theory for wavelet behaviors, the choice of a particular wavelet for a particular problem may even appear arbitrary.  As attempts  to address this, one may find in the literature very many examples of papers---frequently with overlapping content---in which different types of wavelets are applied to the same signal, in order to decide which wavelet is ``better'' in some sense.   The large effort spent on such case-by-case analyses underscores the need for a systematic and unified framework for understanding analytic wavelets and their properties.

Here we show that the generalized Morse wavelets  \cite{daubechies88-ip,bayram00-nnsp,olhede02-itsp,lilly09-itsp} effectively unify all wavelet types mentioned above, as well as the analytic filter and in a sense the complex exponentials themselves, as special cases within a single very broad family.  These wavelets are defined in the frequency domain as
\begin{equation}
\Psi_{\beta,\gamma}(\omega)= \int_{-\infty}^\infty \psi_{\beta,\gamma}(t) e^{-i\omega t}dt= U(\omega) \,a_{\beta,\gamma}\, \omega^\beta e^{-\omega^\gamma}
\label{morse}
\end{equation}
where $a_{\beta,\gamma}$  is a normalization constant, $U(\omega)$ is the unit step function, and $\beta$ and~$\gamma$ are two parameters controlling the wavelet form.\footnote{The generalized Morse wavelet have an additional degree of freedom, the {\em order} $k$, which permits the construction of families of increasingly oscillatory wavelets orthogonal to a chosen wavelet of the form (\ref{morse}), see e.g. \cite{olhede02-itsp}.  Such higher-order wavelets can be expressed as sums of wavelets the form (\ref{morse}) with different $\beta$ and~$\gamma$ values.  This degree of freedom, which is primarily useful in multiple-transform methods [e.g. \citen{bayram00-nnsp}], will not be pursued here.}     The decision of what wavelet to use is greatly simplified by examining the roles of $\beta$ and~$\gamma$ in controlling wavelet properties, as we undertake here, building on recent work \cite{lilly09-itsp}.

\begin{figure}[b!]
        \noindent\begin{center}\includegraphics[width=3.6in]{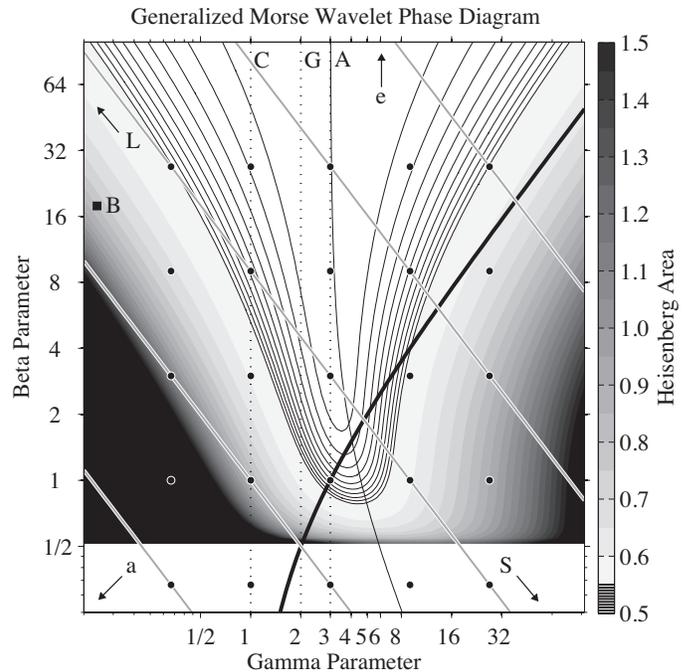}\end{center}
        \caption{\footnotesize        Parameter space for the generalized Morse wavelet superfamily, with the Heisenberg area shaded over $\beta$ and~$\gamma$ on log-log axes.   Note that the Heisenberg area is undefined for $\beta\le 1/2$.      Thin solid lines are contours of the Heisenberg area from 0.5 to 0.55 with a spacing of 0.005.    In this log-log space, curves of constant duration $P_{\beta\gamma}\equiv\sqrt{\beta\gamma}$ are diagonal lines.  Values of $P_{\beta,\gamma}$ of 1/3, 1, 3, 9, and 27 are marked by the thick gray lines.  The dotted vertical lines mark the $\gamma=1$, $\gamma=2$, and $\gamma=3$ wavelet families.  The heavy solid line is the right-hand border to the localization region of \cite{olhede02-itsp}, with the left-hand border being at $\gamma=1$. A thin black line divides wavelets with positive skewness in the frequency domain (on the left) from those with negative skewness (on the right).  Circles mark the locations of wavelets shown in Fig.~\ref{morsies-time}.  Letters mark the locations of other wavelet or analytic filter families, with arrows denoting limits, as discussed in the text: ``L'' for the lognormal wavelets, ``C'' for the Cauchy wavelets, ``G'' for the Derivative of Gaussian wavelets, ``A'' for the Airy wavelets, ``$e$'' for complex exponentials, ``S'' for the Shannon wavelet,  ``a'' for the analytic filter, and ``B'' for the Bessel wavelet. }\label{analytic-morsephase}
\end{figure}\normalsize

\begin{figure*}[t!]
        \noindent\begin{center}\includegraphics[width=3.5in]{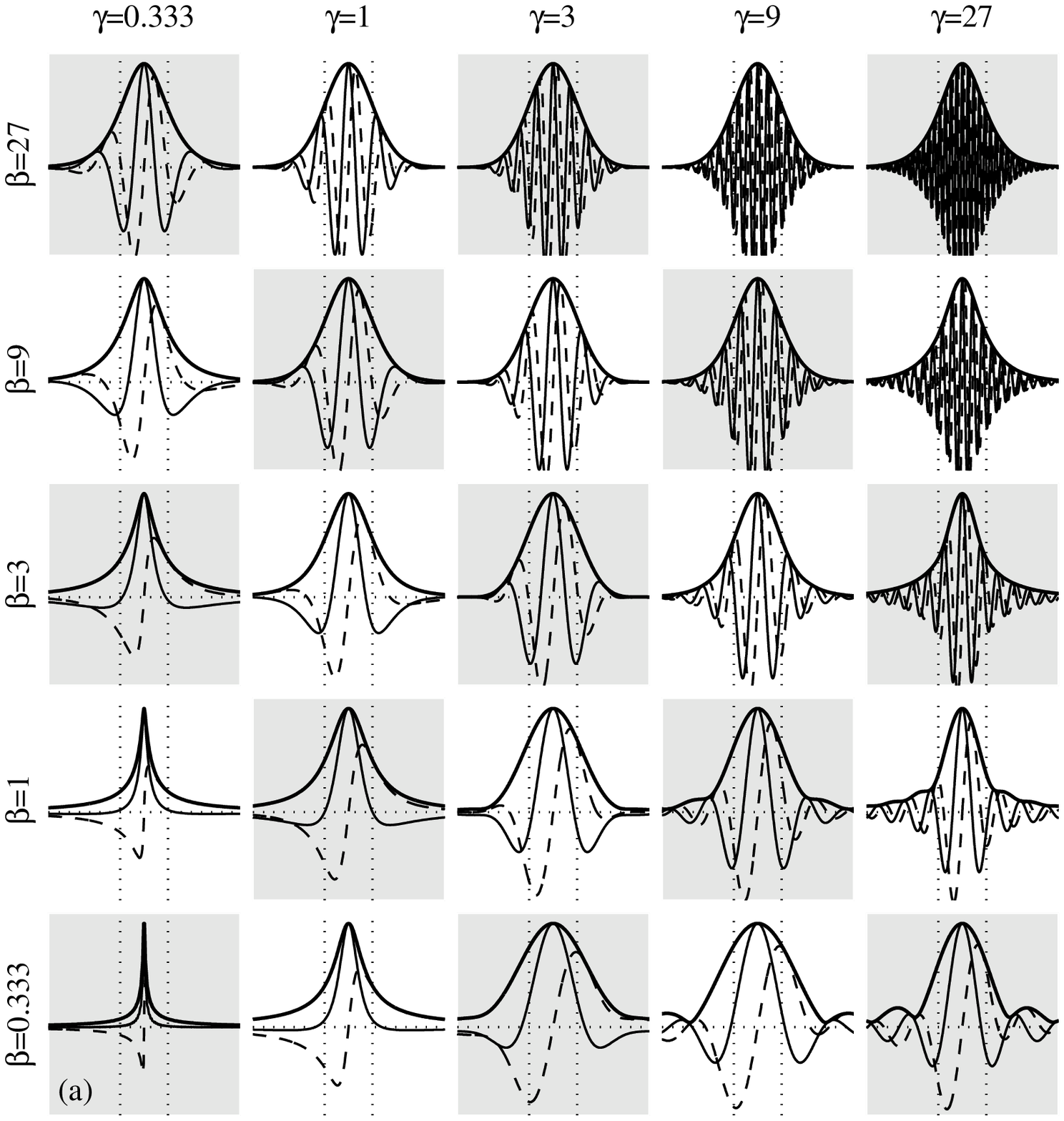}
        \includegraphics[width=3.5in]{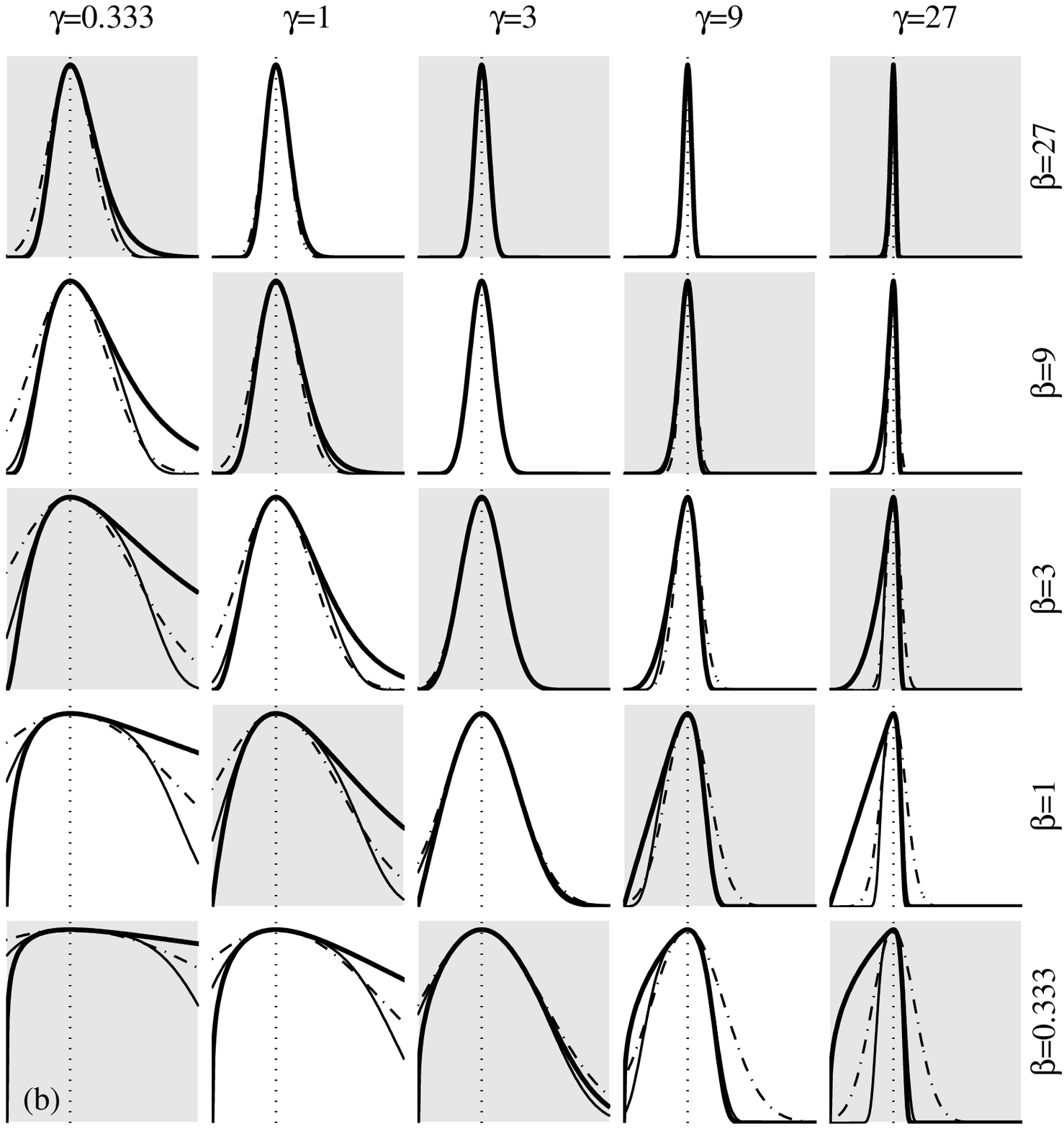}\end{center}
        \caption{\footnotesize
Examples of the generalized Morse wavelets for 25 different $(\beta,\gamma)$ pairs, with time-domain wavelets $\psi_{\beta,\gamma}(t)$ on the left and frequency-domain wavelets $\Psi_{\beta,\gamma}(\omega)$ on the right. In both panels, the parameter ~$\gamma$  increases from left to right, while the parameter $\beta$ increases from  bottom to top.  Both parameters take on the values 1/3, 1, 3, 9, and 27, corresponding to integer powers of 3.  The duration $P_{\beta,\gamma}=\sqrt{\beta\gamma}$ is constant along diagonal lines, as indicated by alternating white and gray backgrounds, and increases from the lower left to the upper right in both panels.  For the time-domain wavelets, a thin solid line is the real part of the wavelet, a thin dashed line is the imaginary part, a heavy solid line is the modulus, a horizontal dotted line marks the zero axis, and vertical dotted lines mark $\pm P_{\beta,\gamma}$.  The time axis has been rescaled so that the duration $P_{\beta,\gamma}$ is the same width in each panel, and the wavelet magnitudes have been rescaled to obtain the same values at time $t=0$.  For the frequency-domain wavelets, a heavy solid line is the wavelet, a thin solid line is the Gaussian approximation, and a thin dash-dotted line is the quartic approximation obtained later from (\ref{freqexpand}) by neglecting the residual term in $f_{\beta,\gamma}(\omega)$.  A vertical dotted line marks the location of the peak frequency $\omega_{\beta,\gamma}$, which always occurs at the same position on account of a rescaling of the frequency axes.
}\label{morsies-time}
\end{figure*}\normalsize

The generalized Morse wavelets were introduced by Daubechies and Paul \cite{daubechies88-ip} as the eigenfunctions of a time/frequency localization operator, later examined in detail by \cite{olhede02-itsp}.  Their name reflects the fact that with  $\gamma=1$, these wavelets become equivalent to a solution to the Schr\"odinger equation examined by Morse \cite{morse29-pr}.  Their frequency-domain form is essentially that of the probability distribution known as a generalized gamma or Stacy distribution \cite{stacy62-ams,lienhard67-qam}.  This distribution has been rediscovered and re-examined multiple times \cite{hegyi99-iwmp}, and finds applications ranging from droplet physics \cite{hegyi99-iwmp} to econometrics \cite{zhang01-jecon} to communications \cite{sagias06-itwc,yacoub07-itvt}.    Consequently, the new results herein and in \cite{lilly09-itsp} regarding the generalized Morse wavelets are also relevant to the use of these functions as probability distributions.

All software associated with this paper is freely distributed to the community as a part of a MATLAB~toolbox, as described in the Appendix.

\section{Generalized Morse Wavelet Parameter Space}

By varying $\beta$ and~$\gamma$, the generalized Morse wavelets can take on a wide variety of forms, in addition to the rescaling by $s$ which dilates or compresses a given wavelet.  A map of generalized Morse wavelet parameter space is presented in Fig.~\ref{analytic-morsephase}.  The dots mark the locations of the illustrative wavelets shown in Fig.~\ref{morsies-time}, in which both time and frequency axes have been rescaled for presentational clarity.   The original derivation of generalized Morse wavelets, see \cite{daubechies88-ip,bayram00-nnsp,olhede02-itsp}, utilized the fact that they are eigenvalues of a time/frequency localization operator for $\gamma\ge 1$ and $\beta>(\gamma-1)/2$, which is the region to the right of the line $\gamma=1$ and to the left of the heavy black curve.  As the generalized Morse wavelets are essentially a probability distribution in the frequency domain, they can be associated with a coefficient of skewness, see eqn. (42) of \cite{lilly09-itsp}.  The thin black line in Fig.~\ref{morsies-time} is the location of vanishing frequency-domain skewness, with positive skewness to the left of this curve and negative skewness to the right.

The shaded quantity in Fig.~\ref{analytic-morsephase} is the Heisenberg area \cite{mallat}, a convenient and conventional measure of time/frequency concentration.  The Heisenberg area of a function $\psi(t)$ is defined as $A_\psi\equiv  \sigma_{t}\,\sigma_{\omega}$, where $\sigma_{t}$ and $\sigma_{\omega}$ are the time-domain and frequency-domain standard deviations, respectively; see e.g. (21) and (22) of \cite{lilly09-itsp} for definitions.  Exact expressions for the  generalized Morse wavelet standard deviations are given in \cite{lilly09-itsp}.    The Heisenberg area $A_{\beta,\gamma}$ of the generalized Morse wavelets decreases with increasing $\beta$ for fixed~$\gamma$. For $\beta\le\frac{1}{2}$, the generalized Morse wavelet temporal standard deviation $\sigma_{t}$ is unbounded, as is the  Heisenberg area.  In the vicinity of the zero-skewness curve, there is a minimum of the Heisenberg area that approaches its theoretical lower bound for the greatest possible degree of concentration at $A_{\beta,\gamma}=\frac{1}{2}$.  With increasing $\beta$ this curve approaches the line $\gamma=3$, corresponding to an important family of  wavelets identified by \cite{lilly09-itsp} as arising from an inhomogeneous Airy function and therefore called the {\em Airy wavelets}.

The wavelets in Fig.~\ref{morsies-time} become more oscillatory in the time domain, and therefore more narrow in the frequency domain, both from left to right as~$\gamma$ increases, and from bottom to top as $\beta$ increases.  The frequency-domain wavelets obtain a maximum value at the {\em peak frequency} $\omega_{\beta,\gamma} \equiv\left(  \beta/\gamma \right) ^{1 /\gamma}$, determined by the frequency at which the derivative of $\Psi_{\beta,\gamma}(\omega)$ with respect to $\omega$ vanishes, see \cite{lilly09-itsp}.    A key parameter is the rescaled second derivative of the frequency-domain wavelet evaluated at its peak frequency,\begin{equation}
P_{\beta,\gamma}^2\equiv-\frac{\omega_{\beta,\gamma}^2\Psi_{\beta,\gamma}''(\omega_{\beta,\gamma})}{\Psi_{\beta,\gamma}(\omega_{\beta,\gamma})} =\beta\gamma.
\end{equation}
It is shown by \cite{lilly09-itsp} that $P_{\beta,\gamma}/\pi$ is the number of oscillations at the peak frequency that fit within the central window of the time-domain wavelet, and therefore $P_{\beta,\gamma}$ is called the wavelet {\em duration}.  $P_{\beta,\gamma}$ is constant along diagonal lines sloping from the upper left the lower right in Figs.~\ref{analytic-morsephase} and~\ref{morsies-time}, so the number of oscillations encompassed by the wavelets along these lines is essentially the same; for example, along the main diagonal in Fig.~\ref{morsies-time}, $P_{\beta,\gamma}=3$ and so the number of oscillations in the central window is $P_{\beta,\gamma}/\pi\approx1$.

Variations of the wavelet form for fixed duration  $P_{\beta,\gamma}=\sqrt{\beta\gamma}$ can be expressed in several ways.  From (\ref{morse}), we see that the parameter $\beta$ controls the behavior near zero frequency, while~$\gamma$  controls the high-frequency decay.  It can be shown \cite{lilly09-itsp} that the wavelet time decay is $\left|\psi_{\beta,\gamma}(t)/\psi_{\beta,\gamma}(0)\right|\sim 1/t^{\beta+1}$, thus  $\beta$ and~$\gamma$  control the time-domain and frequency-domain decay respectively.  Therefore moving along curves of constant $P_{\beta,\gamma}$ from the upper left to the lower right of Fig.~\ref{morsies-time} exchanges a strong long-time decay rate, from $\beta$, for a strong high-frequency decay rate, from~$\gamma$.  Conversely, moving diagonally from the lower left to upper right, the wavelets becomes more oscillatory in the time domain, and more narrow in the frequency domain, in such a way that the ratio of time-domain decay to frequency-domain decay $\beta/\gamma$ remains constant.   It can be seen that the small $\beta$ and large~$\gamma$ wavelets in the lower right-hand corner are problematic on account of undesirable time-domain sidelobes and extreme frequency-domain asymmetry; this is also region within which the generalized Morse wavelet form (\ref{morse}) is no longer the solution to the localization operator problem of \cite{daubechies88-ip,olhede02-itsp}, indicating the practical implications of the operator framework.

Locations of other important wavelets families and analysis functions are indicated in Fig.~\ref{analytic-morsephase} by letters.  Note that herein the  constant $a_{\beta,\gamma}$  in (\ref{morse}) is chosen such that $\Psi_{\beta,\gamma}(\omega_{\beta,\gamma})=2$, as is sensible for the $1/s$ scale normalization in (\ref{vectorwavetrans}) that we prefer, see \cite{lilly09-itsp}.  It was shown by \cite{lilly09-itsp} that the generalized Morse wavelets encompass two other popular families of analytic wavelets: the Morse, also known as the Cauchy or Klauder or Paul, wavelets for the choice $\gamma=1$, and the analytic ``Derivative of Gaussian'' (DoG) wavelets for $\gamma=2$.  Furthermore,  \cite{lilly09-itsp} shows that (\ref{morse}) with $(\beta,\gamma)=\left(0,0\right)$ gives $\psi_{0,0}(t)=\delta(t)+\frac{i}{\pi t}$ or $\Psi_{0,0}(\omega)=2U(\omega)$, which is recognized as twice the analytic filter, while the complex exponentials are approached  in a sense of convergence of moments as $\beta$ approaches infinity for fixed~$\gamma$.  These tendencies are apparent in  Fig.~\ref{morsies-time}: the lower left-hand corner in Fig.~\ref{morsies-time}a resembles an analytic delta-function in the time domain,  while the upper right-hand corner in Fig.~\ref{morsies-time}b resembles a delta-function in frequency.

In Section~\ref{gauss} we will show that with $\gamma=3$, the generalized Morse wavelets closely approximate a Gaussian while remaining exactly analytic, thus implementing the essential idea of the Morlet wavelet without suffering its drawbacks.   In Section~\ref{section:forms} we will compare with three other analytic wavelets, the lognormal or log Gabor \cite{knutsson94-itip,field87-josaa}, Shannon \cite{mallat}, and Bessel \cite{holschneider} wavelets.  The lognormal wavelets are found to be recovered in the limit of vanishing~$\gamma$ and fixed $P_{\beta,\gamma}$, that is, in diagonal motion to the upper left corner, while the Shannon wavelet is recovered in the opposite limit of motion to the lower right corner.  The Bessel wavelet is found to be very closely approximated by the generalized Morse wavelet at $(\beta,\gamma)=\left(22,1/10\right)$.   Altogether this is a fairly exhaustive list of commonly used analytic wavelets and analytic analysis functions, all of which are encompassed by the generalized Morse wavelet form~(\ref{morse}).

\section{Concentration and Gaussianity}\label{gauss}

A desirable characteristic  for a wavelet would be to combine the special properties of a Gaussian---which is in a sense the most time/frequency concentrated function---with the constraints necessary for an analytic wavelet.  This is not achievable; some compromise is necessary, see e.g. \cite{holschneider} p.~36--37.  It will be shown that for integer choices of ~$\gamma$, the $\gamma=3$ family best achieves a high degree of concentration and Gaussianity, while remaining, like all the generalized Morse wavelets, exactly analytic.   The Heisenberg area achieves its theoretical minimum value of one-half for a function which is a Gaussian envelope multiplying a complex exponential, but such a function is not a wavelet because it is not zero mean.  One could say that the idea behind the Morlet wavelet is to introduce a correction in order to recover a zero mean function, hopefully preserving the desirable properties of the Gaussian in the process.  In practice, this works well for oscillatory wavelet settings, but not for time-localized wavelet settings.

The Morlet wavelet is defined in the time domain and the frequency domain as \cite{holschneider,lilly09-itsp}, respectively
\begin{eqnarray}
\psi_{\nu}(t)&=& a_\nu \,  e^{-\frac{1}{2}\,t^2}\left[e^ {i\nu t}-e^ {-\frac{1}{2}\nu^2 }\right]\label{Morletwavelet}\\
\Psi_{\nu}(\omega)&=& a_\nu \,  e^{-\frac{1}{2}\,(\omega-\nu)^2}\left[1-e^ {-\omega\nu}\right]\label{Morletwaveletfrequency}
\end{eqnarray}
where $a_\nu$ is a normalizing constant.  Note that the oscillation frequency $\nu$ is not the same as the peak frequency at which $\Psi_{\nu}'(\omega)$ vanishes; the deviation between these two frequencies grows as $\nu$ become small and vanishes as $\nu$ becomes large, see Appendix~A of \cite{lilly09-itsp}.  For time-localized wavelets with small values of $\nu$,  \cite{lilly09-itsp} shows that the departure of the Morlet wavelet from analyticity---that is,  its frequency-domain support on the negative frequencies---leads to interference terms that significantly degrade the resulting transform.  Here we will show that the correction term also generates deviations from the Gaussian form as well as reductions in the time/frequency localization as measured by the Heisenberg area.

\begin{figure}[t!]
        \noindent\begin{center}\includegraphics[width=3.2in]{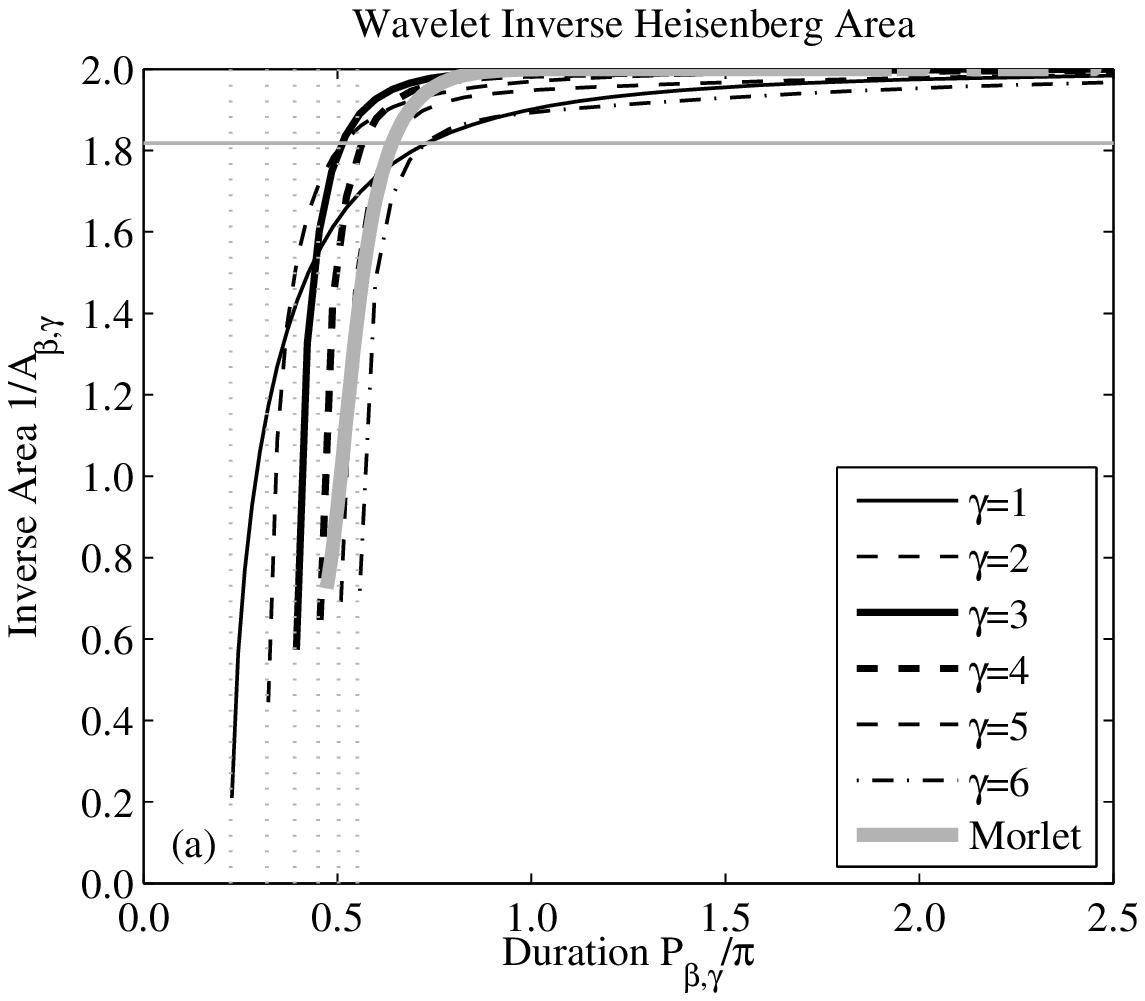}
        \includegraphics[width=3.2in]{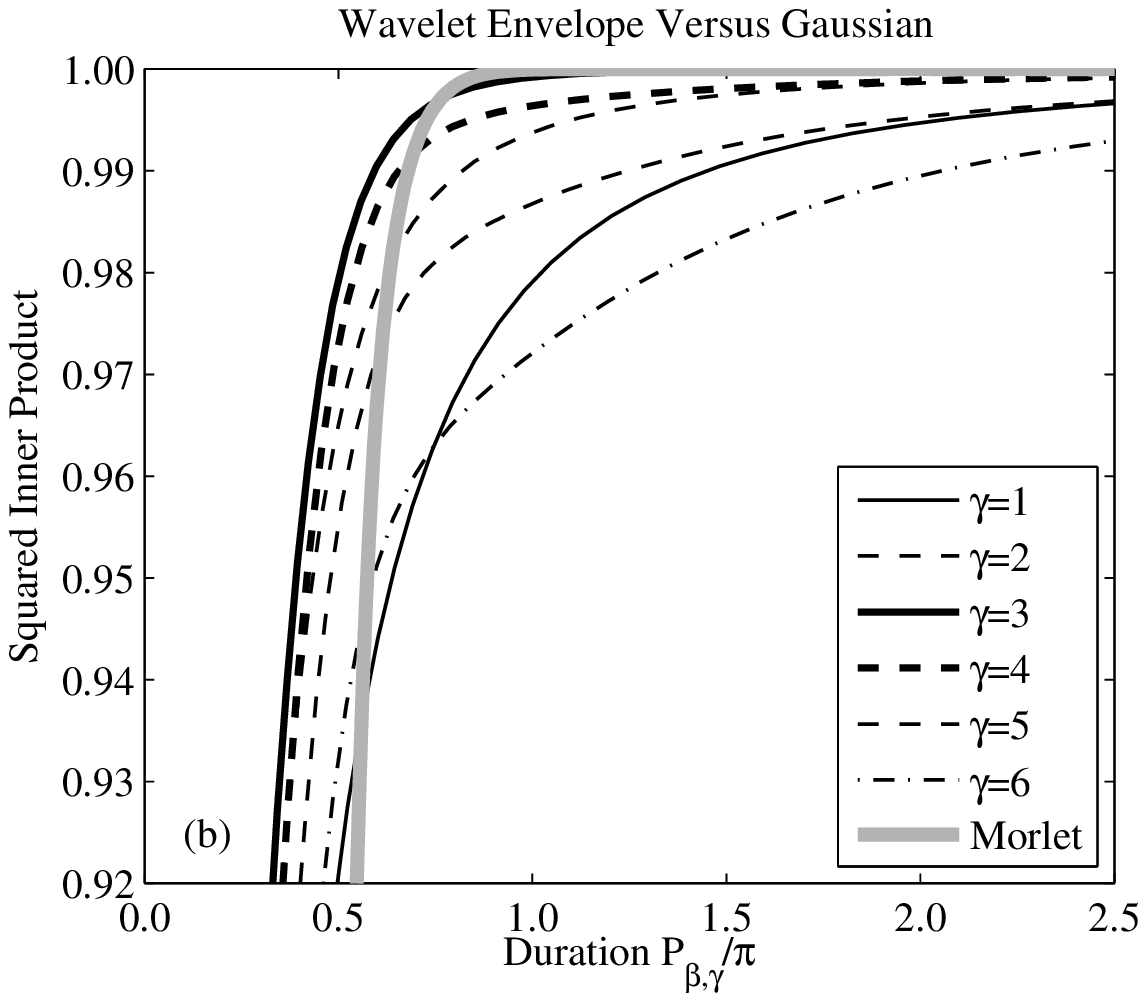}\end{center}
        \caption{\footnotesize (a) The inverse Heisenberg area $1/A_{\beta,\gamma}$ for the generalized Morse wavelets with integer~$\gamma$ from 1 to 6 as a function of wavelet duration $P_{\beta,\gamma}$, as well as for the Morlet wavelet.  Dashed vertical lines marked location at which $\beta=\frac{1}{2}$ for each choice of~$\gamma$, where the Heisenberg area becomes infinite.  The gray horizontal line corresponds to $A_{\beta,\gamma}=0.55$. (b) The squared inner product between the wavelet and the Gaussian approximation
       (\ref{freqexpand}).}\label{morsegaussian}
\end{figure}

The inverse Heisenberg areas of the generalized Morse wavelets families with integer values of~$\gamma$ from one to six are shown in Fig.~\ref{morsegaussian}a, together with the Heisenberg area of the Morlet wavelet.  For large values of the wavelet duration, all wavelet families shown here approach the limiting value of concentration at $A_{\beta,\gamma}=\frac{1}{2}$ or $1/A_{\beta,\gamma}=2$.  For wavelets having small values of the duration, such as one would use to analyze rapidly varying signals, the Morlet wavelet concentration degrades more rapidly as the duration decreases than does the concentration of the $\gamma=2$,~3, or~4 generalized Morse wavelets.  For highly time-localized  settings, the Morlet wavelet become significantly less concentrated than these~$\gamma$-families.

For high degrees of concentration corresponding to $A_\psi\ge 0.55$, the generalized Morse wavelet family for integer~$\gamma$ having the highest degree of time/frequency concentration for fixed choice of duration $P_{\beta,\gamma}$ is the Airy wavelet family at $\gamma=3$.   This appears different from Fig.~\ref{analytic-morsephase}, which shows a minimum in the Heisenberg area somewhat to the right of $\gamma=3$, with the minimum Heisenberg area occurring in the vicinity of $\gamma=4$ for small values of $\beta$ around $\beta\approx 2$.  The difference between these two perspectives is that different~$\gamma$ families should be compared for a fixed value of $P_{\beta,\gamma}$ and not a fixed value of $\beta$.  Taking this into account as in Fig.~\ref{morsegaussian}a shows that the $\gamma=3$ family is the most concentrated family apart from for very small values of the duration, less than about $P_{\beta,\gamma}/\pi\approx \frac{1}{2}$, below which the $\gamma=2$ and then the $\gamma=1$ families become the most concentrated.   The Heisenberg areas of all generalized Morse wavelet families becomes unbounded at $\beta=\frac{1}{2}$, corresponding to $P_{\beta,\gamma}=\sqrt{\gamma/2}$, and thus the smallest possible wavelet duration having a finite area decreases with decreasing~$\gamma$, as indicated by the vertical lines.

\normalsize

A convenient way to express the variation in forms of the generalized Morse wavelets as a function of $\beta$ and~$\gamma$ is via a Fourier-domain expansion about a Gaussian centered at the peak frequency $\omega_{\beta,\gamma}$,
\begin{multline}\label{freqexpand}
\frac{\Psi_{\beta,\gamma}(\omega)}{\Psi_{\beta,\gamma}(\omega_{\beta,\gamma})}= e^{\ln\Psi(\omega)-\ln\Psi(\omega_{\beta,\gamma})}
\\=\exp\left\{-\frac{1}{2}P_{\beta,\gamma}^2 \left(\frac{\omega}{\omega_{\beta,\gamma}}-1\right)^2 + f_{\beta,\gamma}(\omega)\right\}
\end{multline}
where the deviation from Gaussianity is controlled by the function
  \begin{multline} \label{freqexpandetc}
f_{\beta,\gamma}(\omega)=-\frac{1}{6} (\gamma-3)P_{\beta,\gamma}^2\left(\frac{\omega}{\omega_{\beta,\gamma}}-1\right)^3\\-\frac{1}{24}\left[(\gamma-3)^2+2\right] P_{\beta,\gamma}^2\left(\frac{\omega}{\omega_{\beta,\gamma}}-1\right)^4+\Delta\Psi_{\beta,\gamma}^{\{5\}}(\omega).
\end{multline}
This expansion has been accomplished by Taylor-expanding the wavelet logarithm, and making use of expressions for the derivatives of $\ln\Psi_{\beta,\gamma}(\omega)$ from Appendix~D of \cite{lilly09-itsp}.  Here $\Delta\Psi_{\beta,\gamma}^{\{5\}}(\omega)$ is a residual that is defined implicitly  as including the remaining error terms in (\ref{freqexpand}) not explicitly represented in  (\ref{freqexpandetc}).

Observe that in the Gaussian expansion (\ref{freqexpand}), $P_{\beta,\gamma}$ appears as the inverse of the standard deviation of the Gaussian, so that $1/P_{\beta,\gamma}$ is a measure of the frequency-domain bandwidth, while~$\gamma$ controls the degree of deviation from Gaussianity for a fixed choice of $P_{\beta,\gamma}$.  This suggests that $P_{\beta,\gamma}$ and~$\gamma$ are the natural coordinates for the generalized Morse wavelets, with $P_{\beta,\gamma}$ setting the number of oscillations and~$\gamma$ controlling the shape.   When  $\gamma=3$, the lowest-order deviation from Gaussianity---the cubic term---vanishes, while the quartic term obtains its smallest magnitude.   The cubic term controls the {\em local asymmetry} of the wavelet about the peak frequency, while the quartic term describes a local narrowing or broadening of the wavelet peak in comparison with the Gaussian form; both are controlled by~$\gamma$.   Therefore in the sense of the local shape of the Fourier-domain wavelet about the peak frequency $\omega_{\beta,\gamma}$,  the $\gamma=3$ wavelets are the {\em most symmetric} and {\em most nearly Gaussian} members of the generalized Morse wavelet family for a fixed $P_{\beta,\gamma}$ and any integer or non-integer choice of~$\gamma$.

In Fig.~\ref{morsies-time}b, the Fourier-domain wavelets are compared with approximations from keeping only the quadratic (Gaussian) term in (\ref{freqexpand}), or with keeping up to and including the quartic term.  In general, for highly localized wavelets the quartic and even the quadratic approximations provide excellent approximations up to large distances from the peak frequency.  However, for wavelets with a large Heisenberg area and hence poor time/frequency concentration, such as in the lower right-hand corner of Fig.~\ref{morsies-time}b, the low-order Taylor expansion (\ref{freqexpand}) is only accurate in the immediate vicinity of the peak frequency.

A more direct comparison between the wavelets and the Gaussian form is shown in Fig.~\ref{morsegaussian}b.  This is the inner product of the wavelet with the Gaussian form, specifically the quadratic approximation from (\ref{freqexpand}) neglecting the $f_{\beta,
\gamma}(\omega)$ term, after normalizing both functions to unit energy.  For the Morlet wavelet, we use the expressions for the peak frequency and duration given in Appendix~A of \cite{lilly09-itsp}.  Not only is the Morlet wavelet less concentrated than the generalized Morse wavelets, it is also less similar to a Gaussian, despite the intention behind its formation.   While all the wavelets become increasingly more Gaussian as the duration increases, the~$\gamma$-family for integer~$\gamma$ which is always the most similar to a Gaussian for particular choice of $P_{\beta,\gamma}$ is the Airy wavelet family at $\gamma=3$.

\section{Special Forms of Generalized Morse Wavelets}\label{section:forms}

Limiting forms of the generalized Morse wavelets can be found if we rescale the frequency argument $\omega$ by the peak frequency $\omega_{\beta,\gamma}$, defining a new frequency-domain wavelet
\begin{equation}
\Phi_{\beta,\gamma}
\left(\omega\right)\equiv\Psi_{\beta,\gamma}
\left(\omega_{\beta,\gamma}\,\omega \right)
\end{equation}
such that the rescaled wavelet always obtains a maximum  at frequency $\omega=1$. This becomes, from (\ref{morse}),
\begin{equation}
\Phi_{\beta,\gamma}\label{phiexpression}
\left(\omega\right)=2 U(\omega) \,\omega^\beta e^{\beta/\gamma(1-\omega^\gamma)}
\end{equation}
after choosing $a_{\beta,\gamma}$ such that $\Psi_{\beta,\gamma}\left(\omega_{\beta,\gamma}\right)=2$.  Setting $\beta=P^2/\gamma$ with $P$ a positive constant,  this becomes
\begin{equation}
\Phi_{P^2/\gamma,\gamma}
\left(\omega\right)=2 U(\omega) \, e^{ \left(\gamma\ln\omega+1-\omega^\gamma\right) P^2/\gamma^2}
\end{equation}
making use of the fact that $\omega=e^{\ln \omega}$.   The limit of the exponent as~$\gamma$ tends to zero is
\begin{multline}
\lim_{\gamma\,\longrightarrow0} \frac{P^2}{\gamma^2}\left[\gamma\ln \omega+
1-\left(1+\ln \omega \gamma+\frac{1}{2}\gamma^2 \ln^2 \omega+{\cal O}(\gamma^3)\right)\right]\\=-\frac{1}{2}P^2\ln^2\omega
\end{multline}
where we have expanded $\omega^\gamma=e^{\gamma\ln\omega}$ in powers of $\gamma\ln \omega$. Consequently it follows that
\begin{equation}
\lim_{\gamma \rightarrow 0}\Phi_{P^2/\gamma,\gamma}(\omega)=
2U(\omega)e^{-\frac{1}{2}P^2\ln^2(\omega)}\label{lognormal}
\end{equation}
emerges as the small-$\gamma$ limit of the generalized Morse wavelets with $\beta$ chosen such that $P_{\beta,\gamma}=\sqrt{\beta\gamma}$ remains fixed.   The wavelets are lognormal in the frequency domain, with $1/P_{\beta,\gamma}$ playing the role of the standard deviation, and present a  highly asymmetric shape about the maximum at $\omega=1$.  Wavelets of the form (\ref{lognormal}) have been proposed, apparently independently, by \cite{knutsson94-itip} and \cite{field87-josaa}, and appear to be in widespread use under the names ``log Gabor wavelets''  or ``lognormal quadrature wavelets.''

In the opposite limit of infinite~$\gamma$ with $\beta=P^2/\gamma$, moving towards the lower right corner of  Fig.~\ref{analytic-morsephase}, another wavelet emerges.   The $\omega^\beta=\omega^{P^2/\gamma}$ term in (\ref{phiexpression}) tends to unity, while the exponential contains the indeterminate form  $\omega^\gamma/\gamma^2$ which tends to $\frac{1}{2}\,\omega^{\gamma-1}$ by l'H\^ospital's rule.  We then find that (\ref{phiexpression}) becomes
\begin{multline}
\lim_{\gamma\,\longrightarrow\infty}\Phi_{P^2/\gamma,\gamma}
\left(\omega\right)
=\lim_{\gamma\,\longrightarrow\infty} 2\, U(\omega) \,e^{-\frac{1}{2}P^2\omega^{\gamma} } \\ = \left\{\begin{array}{ll} 2  & 0<\omega\le 1\\
                      0, & \omega\le 0, \omega>1  \end{array}\right.  = 2\,\mathrm{rect} \left(\omega-\frac{1}{2}\right)
\end{multline}
where ``$\mathrm{rect}$'' is the unit rectangle function.  Thus the generalized Morse wavelet with~$\gamma$ approaching infinity and $\beta$ approaching zero becomes a bandpass filter with constant amplitude for frequencies $\omega$ between zero and one, and vanishing elsewhere.  This limit is identical with the  definition of the Shannon wavelet $\Psi_{S}(\omega)$ [\citen{mallat},~p.~246], corresponding a sinc function in the time domain
  \begin{equation}
\psi_{S}(t)=\pi\, \sinc\left(\frac{t}{2\pi}\right) \,  e^{i t }\label{Shannonwavelet}.
\end{equation}
The slow $(1/t)$ time decay of this wavelet, however, is an undesirable property that limits its usefulness.

Together the Shannon wavelet and the lognormal wavelets are important because they delimit the most extreme properties that may be obtained for any choice of $P_{\beta,\gamma}$.   Note, however, that these limits are quite different.  As seen in  Fig.~\ref{analytic-morsephase}, the Heisenberg area becomes unbounded in the Shannon limit, but asymptotes to a fixed value for each $P_{\beta,\gamma}$ with decreasing~$\gamma$ in the lognormal limit.  In the Shannon limit the generalized Morse wavelets approach a fixed form for any choice of $P_{\beta,\gamma}$, whereas in the opposite limit of the lognormal wavelets, the wavelet form remains a one-parameter family controlled by $P_{\beta,\gamma}$.  In particular, it is perhaps surprising that the generalized Morse wavelets can both closely approximate a Gaussian in the frequency domain, at $\gamma=3$ as discussed earlier, as well as the lognormal form (\ref{lognormal}).

This means that the only exactly analytic wavelet in common use not included in the generalized Morse wavelet family is the Bessel wavelet  [\citen{holschneider},~p.~32], defined by, with our choice of normalization
\begin{eqnarray}
\psi_{B}(t)&=& \frac{2e^2}{\pi\sqrt{1 -it }}  \, K_1\left(2\sqrt{1 -it }\right) \label{Besselwavelet}\\
\Psi_{B}(\omega)&=&  2e^2e^{-(\omega +1/\omega)}\label{Besselwaveletfrequency}
\end{eqnarray}
where  $K_1(\cdot)$ is the first modified Bessel function of the second kind.   It is evident by comparing (\ref{Besselwaveletfrequency}) and (\ref{morse}) that the Bessel wavelet does not formally lie within the generalized Morse wavelet family.  Nevertheless, a choice of $(\beta,\gamma)$ may be found for which the Bessel wavelet is very closely approximated.  A measure of the similarity between the two wavelets is  the magnitude-squared inner product
\begin{multline}
\alpha^2_{\beta,\gamma}\equiv \\  \left| \int_{-\infty}^\infty\frac{\psi_B(t)}{\sqrt{\int_{-\infty}^\infty \left|\psi_B(u)\right|^2 du}} \frac{\psi_{\beta,\gamma}^*(t)}{\sqrt{\int_{-\infty}^\infty \left|\psi_{\beta,\gamma}(u)\right|^2 du}}\, dt \right|^2.
\end{multline}
A numerical search over a range of $(\beta,\gamma)$ values shows that $\alpha_{\beta,\gamma}^2$ obtains a maximum value of $0.9995$ for $(\beta,\gamma)=\left(22,1/10\right)$.  At this parameter choice, the two time-domain wavelets are visually virtually indistinguishable.

\section{Discussion}\label{section:choice}

This correspondence has shown that a number of commonly used continuous analytic wavelets and analysis filters, rather than being intrinsically distinct, may all be regarded as special cases of a single unified family, the generalized Morse wavelets.  This means that the wavelet transform itself may be regarded as a continuous function of not one but two parameters in addition to time and scale.  The duration or inverse bandwidth $P_{\beta\gamma}=\sqrt{\beta\gamma}$ sets the number of oscillations in the wavelet, while a shape parameter~$\gamma$  shifts the transform between qualitatively different families with the duration held fixed.

The decision of what wavelet to use is greatly simplified by noting the special properties of the Airy wavelet family at $\gamma=3$, previously identified by \cite{lilly09-itsp}.  These are found to be the most Gaussian, most time/frequency concentrated, and most symmetric  member of the generalized Morse wavelet family for any integer~$\gamma$.   We show that the Airy wavelets reflect the spirit behind the Morlet wavelet more so than even the Morlet wavelet itself, and consequently form a natural choice for a default analytic wavelet for general purpose applications.   Particularly if time-localized settings are to be attempted, the Morlet wavelet should be avoided and the Airy wavelet used in its place, reinforcing the conclusion of \cite{lilly09-itsp}.

Of course, real-valued or non-analytic wavelets may be useful, particularly in the analysis of discontinuities \cite{mallat}, but for analyzing oscillatory phenomena analytic wavelets are commonly favored.   There may be occasions when other members of the generalized Morse wavelet family than the Airy wavelets would be more suitable.  One instance arises when the signal of interest may have a form corresponding to one of the wavelet families; this is particular true of the Gaussian and Cauchy families which may provide useful models for time-localized or ``impulsive'' events that themselves resemble wavelets.  Another common scenario is that oscillatory variability may exist close to the Nyquist frequency, in which case one would like to sacrifice symmetry of the wavelet for a more rapid high-frequency decay.   Further investigation of the generalized Morse wavelet parameter space may yield useful information regarding the properties and possible uses of analytic wavelets, in particular, the relationship between the roles of the parameters $\beta$ and~$\gamma$ as differentiation and ``warping'' operators explored by\cite{lilly09-itsp}, and the variations of the wavelet form.

\appendix

\section{A Freely Distributed Software Package}\label{section:software}
All software associated with this paper is distributed as a part of a freely available MATLAB toolbox called JLAB, available at \url{http://www.jmlilly.net}.  The generalized Morse wavelets are implemented with \texttt{morsewave}, while various properties are computed in the functions \texttt{morseprops}, \texttt{morsefreq}, \texttt{morsederiv}, \texttt{morsespace}, and \texttt{morsebox}. The wavelet transform  \texttt{wavetrans} uses the generalized Morse wavelets by default.     Finally, \texttt{makefigs\!\_\,superfamily} generates all figures in this paper.



\begin{thebibliography}{10}
\providecommand{\url}[1]{#1}
\csname url@samestyle\endcsname
\providecommand{\newblock}{\relax}
\providecommand{\bibinfo}[2]{#2}
\providecommand{\BIBentrySTDinterwordspacing}{\spaceskip=0pt\relax}
\providecommand{\BIBentryALTinterwordstretchfactor}{4}
\providecommand{\BIBentryALTinterwordspacing}{\spaceskip=\fontdimen2\font plus
\BIBentryALTinterwordstretchfactor\fontdimen3\font minus
  \fontdimen4\font\relax}
\providecommand{\BIBforeignlanguage}[2]{{%
\expandafter\ifx\csname l@#1\endcsname\relax
\typeout{** WARNING: IEEEtran.bst: No hyphenation pattern has been}%
\typeout{** loaded for the language `#1'. Using the pattern for}%
\typeout{** the default language instead.}%
\else
\language=\csname l@#1\endcsname
\fi
#2}}
\providecommand{\BIBdecl}{\relax}
\BIBdecl

\bibitem{lilly10-itit}
J.~M. Lilly and S.~C. Olhede, ``On the analytic wavelet transform,'' \emph{IEEE
  Trans. Inf. Theory}, vol.~56, no.~8, pp. 4135--4156, 2010.

\bibitem{holschneider}
M.~Holschneider, \emph{Wavelets: an analysis tool}.\hskip 1em plus 0.5em minus
  0.4em\relax Oxford: Oxford University Press, 1995.

\bibitem{mallat}
S.~Mallat, \emph{A wavelet tour of signal processing, 2nd edition}.\hskip 1em
  plus 0.5em minus 0.4em\relax New York: Academic Press, 1999.

\bibitem{knutsson94-itip}
H.~Knutsson, C.-F. Westin, and G.~Granlund, ``Local multiscale frequency and
  bandwidth estimation,'' in \emph{Proceedings of IEEE International Conference
  on Image Processing}.\hskip 1em plus 0.5em minus 0.4em\relax Austin, Texas:
  IEEE, November 1994, pp. 36--40.

\bibitem{field87-josaa}
D.~J. Field, ``Relations between the statistics of natural images and the
  response properties of cortical cells,'' \emph{J. Opt. Soc. Am. A}, vol.~4,
  no.~12, pp. 2379--2394, 1987.

\bibitem{daubechies88-ip}
I.~Daubechies and T.~Paul, ``Time-frequency localisation operators: a geometric
  phase space approach {II}. {T}he use of dilations and translations.''
  \emph{Inverse Probl.}, vol.~4, pp. 661--80, 1988.

\bibitem{bayram00-nnsp}
M.~Bayram and R.~Baraniuk, \emph{Nonlinear and Nonstationary Signal
  Processing}.\hskip 1em plus 0.5em minus 0.4em\relax Cambridge University
  Press, 2000, ch. Multiple window time-varying spectrum estimation, pp.
  292--316.

\bibitem{olhede02-itsp}
S.~C. Olhede and A.~T. Walden, ``Generalized {M}orse wavelets,'' \emph{IEEE
  Trans. Signal Process.}, vol.~50, no.~11, pp. 2661--2670, 2002.

\bibitem{lilly09-itsp}
J.~M. Lilly and S.~C. Olhede, ``Higher-order properties of analytic wavelets,''
  \emph{IEEE Trans. Signal Process.}, vol.~57, no.~1, pp. 146--160, 2009.

\bibitem{morse29-pr}
P.~Morse, ``Diatomic molecules according to the wave mechanics {II}.
  {V}ibrational levels,'' \emph{Physical Review}, vol.~34, pp. 57--64, 1929.

\bibitem{stacy62-ams}
E.~W. Stacy, ``A generalization of the gamma distribution,'' \emph{Ann. Math.
  Stat.}, vol.~33, no.~3, pp. 1187--1192, 1962.

\bibitem{lienhard67-qam}
J.~H. Lienhard and P.~L. Meyer, ``A physical basis for the generalized gamma
  distribution,'' \emph{Quart. Appl. Math.}, vol.~25, no.~3, pp. 330--334,
  1967.

\bibitem{hegyi99-iwmp}
S.~Hegyi, ``A powerful generalization of the {NBD} suggested by {P}eter
  {C}arruthers,'' in \emph{{VIII} International Workshop on Multiparticle
  Production}.\hskip 1em plus 0.5em minus 0.4em\relax Singapore: World
  Scientific Publishing Co. Ltd., 1999, pp. 272--286.

\bibitem{zhang01-jecon}
M.~J. Zhang, J.~R. Russell, and R.~S. Tsay, ``A nonlinear autoregressive
  conditional duration model with applications to financial transaction data,''
  \emph{J. Econometrics}, vol. 104, no. 2001, pp. 179--207, 2001.

\bibitem{sagias06-itwc}
N.~C. Sagias, G.~K. Karagiannidis, P.~T. Mathiopoulos, and T.~A. Tsiftsis, ``On
  the performance analysis of equal-gain diversity receivers over generalized
  gamma fading channels,'' \emph{IEEE Trans. Wirel. Commun.}, vol.~5, no.~10,
  pp. 2967--2975, 2006.

\bibitem{yacoub07-itvt}
M.~D. Yacoub, ``The $\alpha$--$\mu$ distribution: a physical fading model for
  the {S}tacy distribution,'' \emph{IEEE Trans. Veh. Technol.}, vol.~56, no.~1,
  pp. 27--34, 2007.

\end{thebibliography}
\end{document}